\documentclass[12pt]{iopart}
\usepackage[dvips]{graphicx}
\usepackage{cite}
\usepackage{iopams}
\usepackage[dvips]{epsfig}

\newcommand{\be}{\begin{equation}}
\newcommand{\ee}{\end{equation}}
\newcommand{\hs}[1]{\hspace*{#1cm}}
\newcommand{\vs}[1]{\vspace*{#1cm}}
\newcommand{\la}{\langle}
\newcommand{\ra}{\rangle}

\newcommand{\clh}{\mathcal{H}}

\newcommand{\ph}{{\rm ph}}
\newcommand{\Kuchar}{Kucha$\check{{\rm r}}$}
\newcommand{\kuchar}{Kucha$\check{{\rm r}}$ }

\begin{document}

\title{The Conditional Probability Interpretation of the Hamiltonian Constraint.}
  
\author{Carl E. Dolby} \address{Department of Theoretical Physics, 1 Keble Road, 
Oxford, OX1 3RH, UK}

\begin{abstract}
The Conditional Probability Interpretation (CPI), first introduced by Page and 
Wootters \cite{PW}, is reviewed and refined. It is argued that in it's refined form
the CPI is capable of answering various past criticisms \cite{Kuchar,UW,Isham}. 
In particular, questions involving more than one clock time are described in detail, 
resolving the problems raised in \Kuchar's ``reduction ad absurdum'' 
\cite{Kuchar,Page1}. In the case of Parametrized Particle Dynamics, conventional 
quantum mechanics is recovered in the ideal clock limit. When $E=0$ is among 
the continuous spectrum of the Hamiltonian, the 
induced inner product \cite{Rieffel,Landsman,Higuchi,Mar94.1,Mar94.2,Mar99} is used 
to construct the physical Hilbert space $\clh_{\rm ph}$ from the generalized 
eigenvectors in (the topological dual of) $\clh_{\rm aux}$. This allows the CPI 
to be applied to these `continuous-spectrum' cases in a more rigorous fashion than 
that described previously \cite{Page1,Page2,Page3,Page4}. The discrete spectrum case is also treated.
\end{abstract}

\section{Introduction}

 Canonical quantization of General Relativity leads\cite{DeWitt} to the well-known 
`Hamiltonian constraint' 

\be \hat{H} |\psi\ra = 0 \label{1.1} \ee
 (on the {\it physical} Hilbert space $\clh_{\ph}$), known as the Wheeler-DeWitt 
equation. This constraint results from the Dirac-quantization \cite{Dirac,Matschull} 
of any reparametrization-invariant theory (or from the canonical quantization of 
generally covariant theories), and poses immediate problems for the role of time 
is such theories. Combined with the Schr\"{o}dinger equation, it dictates that 
physical states do not change with time! The Heisenberg picture offers no respite, 
since physical operators must satisfy 

\be [\hat{H},\hat{A}] |\psi\ra = 0 \hs{1} \mbox{on} \hs{1} \clh_{\ph} \label{1.2} \ee
 which ensures that all physical properties are constants of motion! Attempts to 
understand this requirement, and to reconcile it with the ever changing evolution 
we see around us, began as soon as the requirement was noticed\cite{DeWitt}, and 
quickly developed into a busy field of investigation. Reviews of this 
topic\cite{Isham,Kuchar} provide a useful chategorisation of the various `problems 
of time', and of the various methods proposed to understand or circumvent them. Among 
those proposals are the so called `timeless interpretations', which accept equations 
(\ref{1.1}), (\ref{1.2}) unaltered, and accept that coordinate time has no 
foundational role to play in the quantum theory. An important task of such 
approaches is to explain how a phenomenological notion of time-evolution can 
emerge in appropriate circumstances. Why, as observers within such a system, 
we can observe changes in the world around us, and importantly, why the changes 
in isolated subsystems can be described by the time-dependent Schr\"{o}dinger equation. 

	It is one such timeless interpretation, the {\it conditional probability 
interpretation} (advanced first by Page and Wootters\cite{PW} and later by 
Page\cite{Page1,Page2,Page3,Page4}) that will be the subject of this article. This 
proposal (herein called CPI) received strong criticism, most notably by \kuchar 
\cite{Kuchar} in his 1992 review article (see also \cite{Isham}). While no proposal 
escaped criticism in that review (with \kuchar concluding that ``In my opinion, none 
of us has so far succeeded in proposing an interpretation of quantum gravity that 
would either solve or circumvent the problems of time'') his criticism of the CPI 
amounted to ``a reduction ad absurdum of the conditional probability interpretation.'' In 
this article we review both the CPI, and the various criticisms of it. We propose a 
refinement of the CPI, and explain how, at least in it's refined form, the CPI is 
capable of answering these criticisms. In particular, it yields reasonable predictions 
when comparing more than one `clock time'.

Section 2 describes briefly how the Hamiltonian constraint emerges in 
reparametrization-invariant particle models, both in classical and quantum 
treatments. We describe how the analogous classical `problem of time' is 
resolved in these models, and lay the groundwork for tackling the quantum 
mechanical case. This is intended as a brief introduction to the concepts 
involved. We stress that later sections are not limited to reparametrization-invariant 
particle models, and apply more generally to systems subject to a Hamiltonian 
constraint. We assume the existence of a Hilbert space $\clh_{\rm tot}$ for the 
unconstrained system, on which the Hamiltonian operator $\hat{H}$ exists and is Hermitian.
In Section 3 we assume also that $E=0$ is among the discrete spectrum of 
$\hat{H}$. We present the refined CPI as it applies to such systems, and explain 
how it differs from the original presentation \cite{PW,Page1}. In particular, 
questions involving more than one `clock time' are described, and \Kuchar's `reduction 
ad absurdum' is addressed and fixed. Section 4 
describes the case when $E=0$ is among the continuous spectrum of the 
Hamiltonian. The `induced' or `spectral analysis' inner product 
\cite{Rieffel,Landsman,Higuchi,Mar94.1,Mar94.2,Mar99}, is used, along with 
the closely related `group averaging' procedure \cite{Ashtekar,MarGi,Mar00} 
to construct the physical Hilbert space $\clh_{\rm ph}$ from the generalized 
eigenvectors in (the topological dual of) $\clh_{\rm tot}$ (only pertinent aspects 
of group averaging and the induced inner product are described in Section 4 - further 
rigor is presented elsewhere \cite{Mar99,Mar00,Landsman,Ashtekar}). This allows the 
CPI to be applied almost unchanged to these `continuous-spectrum' cases. The case of 
Parametrized Particle Dynamics is briefly addressed, and conventional quantum 
mechanics is recovered in the `ideal clock' limit. Conclusions are presented in Section 5. 

\section{Background}

\subsection{Timelessness in Classical Mechanics}

As an example of reparametrization invariant particle dynamics, consider the Jacobi Action
(presented for instance in \cite{LandLif} pg 142, and discussed in detail in \cite{Barbour1,Barbour2}).
\begin{eqnarray} \hs{2} S  & = 2 \int \sqrt{E - V(x^i)} \sqrt{T(\dot{x^i})}  d \lambda \\
 \mbox{where} \hs{1} T  & \equiv \sum_i \frac{m_i}{2} \left(\frac{d x^i}{d \lambda}\right)^2  \end{eqnarray}
	The resulting Lagranges equations are:
\be \frac{d}{d \lambda} \left(\sqrt{\frac{E-V}{T}} m_i \dot{x}^i \right) = - \sqrt{\frac{T}{E-V}}\frac{\partial V}{\partial x^i} \ee
They uniquely determine (for given end points) physical paths in configuration 
space, but leave the parametrization of those paths arbitrary. The transformation 
$x^i(\lambda) \rightarrow x^i(\lambda')$ has the status of a gauge transformation, 
and gauge-invariance leads to the requirement that physical observables be reparametrization 
invariant. Hence physical observables must be {\it constant on physical motions}.

	In the Hamiltonian formulation (see \cite{Dirac,Matschull} for instance) we 
start with the observation that $\frac{\partial L}{\partial \dot{x}^i} \dot{x}^i - L$ is 
identically zero. This leads to the primary (first class) constraint:

\be \phi = \sum_i \frac{p_i^2}{2 m_i} + V(x^i) - E = 0 \label{2.3} \ee
in terms of which the Hamiltonian is simply $H = u \phi$ where $u$ is an arbitrary 
phase space function. Physical observables are then functions $A(p_i,x^i)$ satisfying:
\be \{ A, \phi \} \approx 0 \label{2.8} \ee
where $\approx$ means `equal on the constraint surface'. This is sufficient to ensure that:
\be \{ A, H \} \approx 0 \label{2.5}\ee
and hence that {\it all physical observables are constants of the motion}. We note 
in passing that not all physical observables $A \neq B$ are distinct, since they 
may satisfy $A \approx B$, in which case they represent the same physical observable.

	We see that even in the classical description of reparametrization invariant 
systems, physical observables have no dependence on `coordinate time' $\lambda$. To 
understand how change can still be described in these timeless situations, consider a 
physical path $\gamma$ in configuration space, and consider questions such as:

\begin{itemize}

\item Does the path pass through a given region $V \in Q$? 

\item Does $x^2 = X$ anywhere on $\gamma$? If so, what is $x^1$ when $x^2 = X$?

\end{itemize}

These are well-posed questions, the answers to which are independent of the 
parametrization of $\gamma$. In the second question, we could consider how the 
value of $x^1$ changes as we vary $X$. By so doing we can describe how properties 
of the system ($x^1$ here) change w.r.t. other properties ($x^2$ here) of the 
system. Sufficiently large complex systems will typically include many 
subsystems (clocks, planetary motions etc) which could be used to describe 
changes in other subsystems in essentially this way. In laboratory physics for 
instance, we measure how a chosen (sub)subsystem changes (`evolves') w.r.t. changes 
in {\it clock time}. Since the clock is part of the total system, then this change 
is just like the second question above. It depends only on the path in configuration 
space (which includes the clocks configuration), and not on the parametrization of 
that path. It is this kind of change, which does not require `coordinate time' 
$\lambda$, which we seek to describe in the quantized theory.

	Another simple example of a reparametrization invariant system is provided 
by `Parametrized Particle Dynamics'. In this case we start with a (non-relativistic) 
Lagrangian, such as $\tilde{L}(x^i,\frac{d x^i}{d t}) = \sum_{i} m_i \left(\frac{d 
x^i}{d t}\right)^2  - V(x^i)$. By introducing an arbitrary parameter $\lambda$ we 
can write the action as:

\begin{eqnarray} S & = \int {\rm d} \lambda \, L(x^i,t,\dot{x},\dot{t}) \\
\hs{-3} \mbox{ where } L(x^i,t,\dot{x},\dot{t}) & \equiv \dot{t} \tilde{L}(x^i,\frac{\dot{x}^i}{\dot{t}^i}) \hs{1} \mbox{and } \dot{} \equiv \frac{d}{d \lambda} \end{eqnarray}

The coordinate $t$ is now part of the configuration space. It must represent an observable property of the system - a `clock variable'. The action is reparametrization invariant w.r.t. the independent variable $\lambda$, which gives the Hamiltonian constraint:
\be H_{\rm tot} = p_t + \tilde{H}(x^i,p_i) \approx 0 \label{PPD1} \ee
where $\tilde{H}(x^i,p_i)$ is the Hamiltonian appropriate to the original Lagrangian. Equation (\ref{2.5}) requires that physical observables be independent of $\lambda$, but allows dependence on $t$. Equation (\ref{PPD1}) ensures that the dependence on $t$ agrees with that ascribed by $\tilde{H}(x^i,p_i)$.

\subsection{Quantizing Reparametrization-Invariant Systems.}

The quantization of constrained systems is described in various texts\cite{Dirac,Matschull,Ashtekar2}. Our treatment is brief, and covers only salient features:

\begin{enumerate}

\item Find a mapping $F \rightarrow \hat{F}$ from functions on phase space to 
operators on some vector space $\clh_{{\rm aux}}$, satisfying $[\hat{F},\hat{G}] = i 
\hbar \widehat{\{F,G\}} + O(\hbar^2)$. 

\item The constraint $H \approx 0$, is replaced by the operator equation 
$\hat{H} |\psi\ra = 0$, which defines a subspace $\clh_{{\rm ph}}$ of 
$\clh_{{\rm aux}}$ - the space of {\it physical states}. Equation (\ref{2.8}) 
leads to equation (\ref{1.2}), and hence to the requirement that physical operators 
map physical states onto physical states. We assume for convenience that there is 
only one constraint (the Hamiltonian constraint) left to be solved in 
obtaining $\clh_{\rm ph}$. If instead there is a family 
$\hat{C}_i$ of constraint operators then, provided they generate a unitary group 
\cite{Ashtekar}, a minor generalization of our treatment can still apply. 

\item An inner product is defined on $\clh_{{\rm ph}}$ such that real observables 
$A$ map onto Hermitian operators $\hat{A}_{{\rm ph}}$ on $\clh_{{\rm ph}}$. This 
normally allows a unitary evolution to be defined on $\clh_{{\rm ph}}$ by the 
well-known Schr\"{o}dinger equation $i \hbar \frac{d}{d \lambda} |\psi \ra = \hat{H} 
|\psi \ra$, in terms of which the standard probabilistic interpretation follows. Combined 
with step (ii) however, this just tells us that physical states do not evolve with 
respect to $\lambda$. The relevant probabilistic interpretation of the theory is 
then more subtle, and is described in the next section.

\end{enumerate}

Notice that step (iii) only requires $\clh_{\rm ph}$ (and not $\clh_{\rm aux}$) 
to be Hilbert. We will assume 
however (as is commonly the case) that the unconstrained space $\clh_{\rm aux}$ is 
equipped with an inner product with respect to which $\hat{H}$ is Hermitian. This 
is the case for instance in the `connection representation' of quantum gravity 
\cite{Ashtekar,Ashtekar2} and in Loop Quantum Gravity \cite{Rovelli,Rovelli2}, where 
a Hilbert space $\clh_{\rm aux}$ has been found on which the Gauss and diffeomorphism 
constraints have been solved, and only the Hamiltonian constraint remains to be 
rigourously solved. (While the Hamiltonian constraint in those theories depends 
non-trivially on an arbitrary smearing function $N(x)$, Thiemann has recently 
proposed a resolution of this difficulty, by replacing the constraints with 
one unique `Master Constraint' $\hat{M}$ \cite{Thiemann}.) 

For simplicity we also assume in Section 3 that the constraint equation specifies 
a `bound' eigenspace, in the sense that $E=0$ is among the discrete spectrum of 
$\hat{H}$. In Section 4 we describe how the CPI can be applied also when $E=0$ is 
among the continuous spectrum of $\hat{H}$.

\section{The Refined CPI for the `Bound' Case}

If $E=0$ is among the discrete spectrum of $\hat{H}$, then $\clh_{{\rm ph}}$ is 
a subspace of $\clh_{{\rm aux}}$; the inner product on $\clh_{\rm ph}$ is the same as 
that on $\clh_{\rm aux}$. The projection operator $\hat{P}^{{\rm ph}}: \clh_{{\rm aux}}
\rightarrow \clh_{{\rm ph}}$ can be used to generate a physical state $|\psi_{{\rm ph}}\ra 
\equiv \hat{P}^{{\rm ph}} |\psi\ra$ from any state $|\psi\ra \in \clh_{{\rm aux}}$ and 
to generate a physical operator:

\be \hat{A}^{{\rm ph}} \equiv \hat{P}^{{\rm ph}} \hat{A} \hat{P}^{{\rm ph}} \ee

from any operator $\hat{A}$ on $\clh_{{\rm aux}}$. However, distinct operators 
$\hat{A},\hat{B}$ on $\clh_{{\rm aux}}$ only generate distinct physical properties 
if $\hat{A}^{{\rm ph}} \neq \hat{B}^{{\rm ph}}$. Projection operators $\hat{P}_{A}$ on 
$\clh_{{\rm aux}}$ generate POVM's $\hat{P}^{{\rm ph}}_{A}$ on $\clh_{{\rm ph}}$. 

	It is sometimes convenient to write the projection operator $\hat{P}^{{\rm ph}}$ as:
\be \hat{P}^{{\rm ph}} | \psi \ra = \lim_{\tau \rightarrow \infty} \frac{1}{2 \tau} \int_{-\tau}^{\tau} {\rm d} a \,  e^{-i \hat{H} a} |\psi\ra \label{13} \ee

which is related to the process of `group averaging' often used to provide a 
formal solution of the constraint equation \cite{Ashtekar,Mar00} (see also 
\cite{Rovelli2} for a similar construction in Loop Quantum Gravity), as will be 
described further in Section 4. Equation (\ref{13}) is easily verified by expanding 
$|\psi\ra$ in terms of the spectrum of $\hat{H}$. While physical states have no 
dependence on coordinate time, equation (\ref{13}) shows that they can be written 
as the time-average of Schr\"{o}dinger-evolved states. In this sense the physical 
state $| \psi \ra$ is associated with an entire history of the system, rather than 
with any point on that history. It is the quantum analog of the `path in configuration 
space' discussed in the last section. The integration variable $a$ in 
equation (\ref{13}) is nothing more than a variable of integration. (While it is 
sometimes convenient to introduce a `lapse 
function' $N(\lambda)$ and put ${\rm d} a = N(\lambda) d \lambda$ \cite{Mar95}, this 
is not necessary.) Physical density operators can similarly be written as:

\be \hat{\rho}^{{\rm ph}} \equiv \hat{P}^{{\rm ph}} \hat{\rho} \hat{P}^{{\rm ph}}
 = \lim_{\tau \rightarrow \infty} \frac{1}{(2 \tau)^2} \int_{-\tau}^{\tau} {\rm d} a \int_{-\tau}^{\tau} {\rm d} a' e^{-i \hat{H} a} \hat{\rho} e^{-i \hat{H} a'} \label{2.10}\ee

	Given a projection operator $\hat{P}_A$ on $\clh_{{\rm aux}}$ and 
a physical density operator $\hat{\rho}^{{\rm ph}}$ we identify 
\be P(A; \hat{\rho}) \equiv \frac{\Tr(\hat{P}^{{\rm ph}}_{A} \hat{\rho}^{{\rm ph}})}{\Tr(\hat{\rho}^{{\rm ph}})} = \frac{\Tr(\hat{P}^{{\rm ph}} \hat{P}_{A} \hat{P}^{{\rm ph}} \hat{\rho})}{\Tr(\hat{P}^{{\rm ph}} \hat{\rho})} \label{3.1} \ee

with the apriori probability of $A$ in $\hat{\rho}^{\rm ph}$. This does not 
refer to the probability of $A$ 
`at some given time', and at this stage bears little resemblance to the probabilities 
of our intuition. In semiclassical terms, it is perhaps best thought of as 
representing the `proportion of the physical path (on configuration space) on which
$A$ is true'. While we can replace $\hat{P}^{{\rm ph}}_{A}$ with $\hat{P}_{A}$ 
in (\ref{3.1}) without altering the result, we should not forget that it is the 
physical operator $\hat{P}^{{\rm ph}}_{A}$ that we are measuring. Similarly, we 
could replace $\hat{\rho}^{{\rm ph}}$ with $\hat{\rho}$ in equation (\ref{3.1}), 
without change. (We cannot replace both $\hat{P}^{{\rm ph}}_{A}$ and 
$\hat{\rho}^{{\rm ph}}$ by their unphysical counterparts.)

Given two commuting projection operators $\hat{P}_A$ and $\hat{P}_B$ on $\clh_{\rm aux}$ 
we can define $P(A \mbox{ when } B ; \hat{\rho})$ in terms of $\clh_{{\rm ph}}$ by:

\begin{eqnarray} P(A \mbox{ when } B ; \hat{\rho}) & \equiv \frac{\Tr(\hat{P}^{{\rm ph}} 
\hat{P}_{A} \hat{P}_{B} \hat{P}^{{\rm ph}} \hat{\rho}^{{\rm ph}})}{\Tr(\hat{P}^{{\rm ph}} 
\hat{P}_{B} \hat{P}^{{\rm ph}} \hat{\rho}^{{\rm ph}})} \, = \frac{\Tr(\hat{P}^{{\rm ph}}_{AB} 
\hat{\rho}^{{\rm ph}})}{\Tr(\hat{P}^{{\rm ph}}_B \hat{\rho}^{{\rm ph}})} 
\label{3.3} \end{eqnarray}

It can be written in terms of measurement operators (see for instance 
\cite{Nielson} pg 85 and 102) as:

\be P(A \mbox{ when } B ; \hat{\rho}) = \frac{\Tr(\hat{M}^{\dagger}_{AB} 
\hat{M}_{AB} \hat{\rho}^{{\rm ph}})}{\Tr(\hat{M}^{\dagger}_{B} \hat{M}_{B} 
\hat{\rho}^{{\rm ph}})} = \frac{\Tr(\hat{M}^{\dagger}_{AB} \hat{M}_{AB} 
\hat{\rho}^{{\rm ph}})}{\sum_A \Tr(\hat{M}^{\dagger}_{AB} \hat{M}_{AB} 
\hat{\rho}^{{\rm ph}})} \ee
where $\hat{M}_{AB} = \hat{P}_A \hat{P}_B \hat{P}^{{\rm ph}}$, $\hat{M}_A = 
\hat{P}_A \hat{P}^{{\rm ph}}$ and 
$\hat{M}_B = \hat{P}_B \hat{P}^{{\rm ph}}$. These are measurement operators 
(in the sense of \cite{Nielson}) but are not projective 
measurements. We will call operators of the form $\hat{M}_X = \hat{P}_X \hat{P}^{{\rm ph}}$ 
``physical measurement operators'', since $\hat{M}^{\dagger}_X \hat{M}_X = 
\hat{P}_X^{{\rm ph}}$ is a physical operator. 

	$P(A \mbox{ when } B ; \hat{\rho})$ is best thought of as representing the 
proprtion of the physical path on which $A$ and $B$ are simultaneously true, divided by 
the proportion on which $B$ is true. It is {\it not} the same as the probability of $A$ 
{\it given} $B$, $P(A|B;\hat{\rho})$, which is instead given by:

\begin{eqnarray} P(A|B ; \hat{\rho}) & = \frac{\Tr(\hat{P}^{{\rm ph}}_{A} \hat{P}_{B}^{{\rm ph}} \hat{\rho}^{{\rm ph}} \hat{P}^{{\rm ph}}_{B})}{\Tr(\hat{P}^{{\rm ph}}_B \hat{\rho}^{{\rm ph}} \hat{P}^{{\rm ph}}_{B})} \, = \frac{\Tr(\hat{M}^{\dagger}_B \hat{M}^{\dagger}_A \hat{M}_A \hat{M}_B \hat{\rho}^{{\rm ph}})}{\sum_A \Tr( \hat{M}^{\dagger}_B \hat{M}^{\dagger}_A \hat{M}_A \hat{M}_B \hat{\rho}^{{\rm ph}})}  \label{3.4} \\
 & = P(A ; \hat{\rho}_B) \hs{1} \mbox{where}  \hs{1} 
\hat{\rho}_B \equiv \frac{\hat{M}_B \hat{\rho}^{{\rm ph}} \hat{M}^{\dagger}_B}{\Tr( \hat{M}^{\dagger}_B \hat{M}_B \hat{\rho}^{{\rm ph}})} \label{3.5} \end{eqnarray}

Although $\hat{\rho}_B$ in 
equation (\ref{3.5}) is unphysical, both $P(A|B ; \hat{\rho})$ and 
$P(B ; \hat{\rho})$ are physical operations, since $\hat{M}^{\dagger}_A \hat{M}_A$ 
and $\hat{M}^{\dagger}_B \hat{M}_B$ are physical. (Replacing $\hat{\rho}_B$ with 
$\hat{\rho}^{\rm ph}_B$ does not alter the result.) To appreciate the logical 
distinction between $P(A \mbox{ when } B ; \hat{\rho})$ and $P(A|B ; \hat{\rho})$ recall 
that a physical state $|\psi \ra$ here is the quantum mechanical equivalent of a path 
in configuration space, not of a point on it. The question ``does a path pass through volume $V_2$ {\it given} that it passes through volume $V_1$?'' is clearly distinct from the question ``does a path pass through volume $V_2$ {\it when} it passes through volume $V_1$?''. 

	In Pages original writings \cite{PW,Page1,Page2,Page3,Page4},  
equation (\ref{3.3}) was referred to as the conditional probability $P(A|B)$, 
despite acknowledging it's distinction from equation (\ref{3.4}). We consider that to 
have been misleading, and have rectified it here. Equations (\ref{3.3}) and 
(\ref{3.5}) can also be combined, to give for instance:

\begin{eqnarray}
\hs{-2} P(A_2 \mbox{ when } B_2 | A_1 & \mbox{ when } B_1 ; \hat{\rho}) = P(A_2 \mbox{ when } 
B_2; \hat{\rho}_{A_1 B_1}) \label{twotime1} \\
& \hs{3} \mbox{ where } \hat{\rho}_{A_1 B_1} = \frac{\hat{M}_{A_1 B_1} 
\hat{\rho}^{\rm ph} \hat{M}^{\dagger}_{A_1 B_1}}{\Tr(\hat{M}^{\dagger}_{A_1 B_1} 
\hat{M}_{A_1 B_1} \hat{\rho}^{\rm ph})} \\
 & = \frac{\Tr(\hat{P}^{{\rm ph}}_{A_1 B_1} \hat{P}^{{\rm ph}}_{A_2 B_2} 
\hat{P}_{A_1 B_1}^{{\rm ph}} \hat{\rho}^{{\rm ph}})}{\Tr(\hat{P}^{{\rm ph}}_{A_1 B_1} 
\hat{P}^{{\rm ph}}_{B_2} \hat{P}_{A_1 B_1}^{{\rm ph}} \hat{\rho}^{{\rm ph}})} \label{2times}\\
 & = \frac{\Tr(\hat{P}^{{\rm ph}}_{A_1 B_1} \hat{P}^{{\rm ph}}_{A_2 B_2} 
\hat{P}_{A_1 B_1}^{{\rm ph}} \hat{\rho}^{{\rm ph}})}{\sum_A \Tr(\hat{P}^{{\rm ph}}_{A_1 B_1} 
\hat{P}^{{\rm ph}}_{A B_2} \hat{P}_{A_1 B_1}^{{\rm ph}} \hat{\rho}^{{\rm ph}})} \label{twotime2} \end{eqnarray}

 We will return to this expression later. For now, we consider equation (\ref{3.3}) for 
$P(A \mbox{ when } B ;\hat{\rho})$. Our interpretation of equation (\ref{3.3}) is 
justified simply because, as will now be shown, it accords with our 
conventional notion of `when' as being `at the same time as', whenever a 
reliable `time' can be defined. For this purpose, suppose now that the total 
Hamiltonian can be written as:
\be \hat{H} \approx \hat{H}_{s} + \hat{H}_{c} \label{3.7} \ee
where $\hat{H}_c$ describes a `clock Hamiltonian', $\hat{H}_s$ a `system 
Hamiltonian', and $[\hat{H}_s,\hat{H}_c] = 0$. Then $\clh_{{\rm aux}}$ can be 
decomposed as $\clh_c \otimes \clh_s$, and we can seek a 1-parameter family of 
clock projection operators $\hat{P}_T$ on $\clh_c$ satisfying:

\be \hat{P}_T \approx e^{-i \hat{H}_c T} \hat{P}_0 e^{i \hat{H}_c T}  \label{3.8} \ee

We could require exact equality in equation (\ref{3.8}) and use it to define 
$\hat{P}_T$ in terms of $\hat{P}_0$, as was done for instance in \cite{PW}. However 
the more general condition (\ref{3.8}) is still sufficient, and we note that it need 
only be defined for a suitable finite range of the parameter $T$ (the `clocks lifetime' 
for instance). Given an operator $\hat{P}_A$ on $\clh_s$ we can now calculate

\begin{eqnarray}
P(A \mbox{ when } T;\hat{\rho}) & = \frac{\Tr_{{\rm ph}}(\hat{P}^{{\rm ph}} \hat{P}_{A} \hat{P}_{T} \hat{P}^{{\rm ph}} \hat{\rho}^{{\rm ph}})}{\Tr_{{\rm ph}}(\hat{P}^{{\rm ph}} \hat{P}_{T} \hat{P}^{{\rm ph}} \hat{\rho}^{{\rm ph}})} \, 
= \frac{\Tr_{{\rm aux}}(\hat{P}_{A} \hat{P}_{T} \hat{\rho}^{{\rm ph}})}{\Tr_{{\rm aux}}(\hat{P}_{T} \hat{\rho}^{{\rm ph}})} \label{essence1}\\
 & \approx \frac{\Tr_{{\rm aux}}(\hat{P}_{A} \hat{P}_{0} e^{i \hat{H}_c T} \hat{\rho}^{{\rm ph}} e^{- i \hat{H}_c T})}{\Tr_{{\rm aux}}(\hat{P}_{0} e^{i \hat{H}_c T} \hat{\rho}^{{\rm ph}}e^{-i \hat{H}_c T})} \\
 & \approx \frac{\Tr_{{\rm aux}}(\hat{P}_{A} \hat{P}_{0} e^{-i \hat{H}_s T} \hat{\rho}^{{\rm ph}} e^{ i \hat{H}_s T})}{\Tr_{{\rm aux}}(\hat{P}_{0} e^{-i \hat{H}_s T} \hat{\rho}^{{\rm ph}}e^{i \hat{H}_s T})} \label{essence2} \\
 & = \Tr_{{\rm s}}(\hat{P}_{A} e^{-i \hat{H}_s T} \hat{\rho}^{{\rm s}}e^{i \hat{H}_s T}) \hs{.3} \mbox{ where } \hs{.5} \hat{\rho}^{{\rm s}} \equiv \frac{\Tr_{{\rm c}}( \hat{P}_{0} \hat{\rho}^{{\rm ph}})}{\Tr_{{\rm aux}}(\hat{P}_{0} \hat{\rho}^{{\rm ph}})} \label{3.11} 
\end{eqnarray}

	That is, the probability of $A$ at clock time $T$ behaves {\it as if} we 
Schr\"{o}dinger-evolved the `system state' $\hat{\rho}^{{\rm s}}$ in $\clh_s$. This is 
the essence of the Conditional Probability Interpretation\footnote{The 
original papers \cite{PW,Page4} dealt with closed systems generally, and so 
used the density operator $\bar{\rho} = \lim_{T \rightarrow \infty} \frac{1}{2 T} 
\int_{-T}^{T} {\rm d} t \, e^{-i \hat{H} t} \, \hat{\rho} \, e^{ i \hat{H} t}$ rather 
than equation (\ref{2.10}). It was acknowledged already in \cite{Page4} that 
this should be suitably modified when considering the Hamiltonian constraint.}. 

	It was emphasized by Kuchar \cite{Kuchar}, that neither the operator 
$\hat{P}_A$ nor the density operator $\hat{\rho}_s$ is physical, in the sense of equations 
(\ref{1.1}) and (\ref{1.2}). This is certainly true,  but unimportant, since 
$\hat{P}_{AT}^{{\rm ph}}$ and $\rho^{{\rm ph}}$ are physical, so $P(A \mbox{ when } T)$ 
is undeniably physical. It is necessary for consistency to demonstrate that this 
physical probability behaves {\it as if} we Schr\"{o}dinger-evolved 
$\hat{\rho}^{{\rm s}}$ in $\clh_s$, but we needn't accept the separate reality of 
$\hat{\rho}^{{\rm s}}$ or of $P_A$ in order to achieve this.

	In equations (\ref{essence1}) - (\ref{3.11})  no requirement was placed on 
the commutator of clock 
operators denoting different `clock times'. $[\hat{P}_{T_1},\hat{P}_{T_2}] \neq 0$ 
in general, so different clock states will not distinguish `different times' 
with absolute certainty. A desirable property for a `good clock' is that 
$\Tr_c(\hat{P}_{T_1}\hat{P}_{T_2})$ be sufficiently small whenever $|T_1 - T_2|$ is 
sufficiently large. A choice of `good clock' will depend in general on what 
subsystem $\clh_s$ we are probing, and on what accuracy we desire. Most choices of 
$\hat{H}_c$ will not admit `perfect clocks' (satisfying $\hat{P}_{T_1}\hat{P}_{T_2} = 
\delta(T_1 - T_2)$ ), although one exception is Parametrized Particle dynamics, which 
we consider briefly in Section 4.  It is a virtue of the CPI that it does not require a 
`perfect clock'. Nor does it require an `internal time variable' $T(x,p)$ on phase 
space, or a `time operator' $\hat{T}$ on $\clh_{{\rm aux}}$ (contrary to beliefs 
expressed elsewhere \cite{UW,Isham}).

	The state $\hat{\rho}^s(T) = e^{-i \hat{H}_s T} \hat{\rho}^s e^{i \hat{H}_s T}$ 
in equation (\ref{3.11}) is 
to be interpreted as representing the state of the system when the clock reads $T$. To 
investigate this more fully, suppose that equations (\ref{3.7}) and (\ref{3.8}) are strict 
equalities, and that 
\be \hat{P}_T = |\psi_c(T)\ra \la \psi_c(T)| \label{new30} \ee
where $|\psi_c(T)\ra \equiv e^{-i \hat{H}_c T} |\psi_c\ra$ in $\clh_c$. For convenience, suppose further that $\hat{\rho}^{{\rm ph}}$ is given by:

\be \hat{\rho}^{{\rm ph}} = \hat{P}^{{\rm ph}} |\psi_s\ra |\psi_c\ra \la \psi_c|\la \psi_s| \hat{P}^{{\rm ph}} \label{new31}\ee
where $|\psi_s\ra \in \clh_s$ is a chosen `initial system state'. Equation (\ref{13}) can be used to write the `effective system density operator' $\hat{\rho}^s(T)$ on $\clh_s$ as

\be \hat{\rho}^s(T) = \frac{|\psi^{{\rm eff}}_s(T)\ra \la \psi^{{\rm eff}}_s(T) |}{\la \psi^{{\rm eff}}_s(T)|\psi^{{\rm eff}}_s(T)\ra} \label{3.15} \ee
where the `effective system state' $|\psi^{{\rm eff}}_s(T)\ra$ is given by

\begin{eqnarray} |\psi^{{\rm eff}}_s(T)\ra & = \lim_{\tau \rightarrow \infty}  \frac{1}{2 \tau} \int_{-\tau}^{\tau} {\rm d} a \, f_c(a) e^{- i \hat{H}_s (T+a)} |\psi_s\ra = e^{- i \hat{H}_s T} |\psi^{{\rm eff}}_s\ra \label{3.10} \\
\hs{-1} \mbox{ where } \hs{.2} f_c(a) & \equiv \la \psi_c | \psi_c(a)\ra = \la \psi_c |e^{-i \hat{H}_c a} |\psi_c\ra \\
\hs{-1} \mbox{ and } \hs{.4} |\psi^{{\rm eff}}_s\ra & \equiv \lim_{\tau \rightarrow \infty}  \frac{1}{2 \tau} \int_{-\tau}^{\tau} {\rm d} a \, f_c(a) e^{- i \hat{H}_s a} |\psi_s\ra \end{eqnarray}

Notice that $|f_c(a)|^2 = \Tr_c(\hat{P}_0 \hat{P}_a) = \Tr_c(\hat{P}_T \hat{P}_{T+a})$ provides a 
measure of the clocks effectiveness at distinguishing different `clock times'. 
If $|\psi_c\ra$ was a `good clock' then $f_c(a)$ would be sharply peaked about 
$a=0$ and we would have (up to scale) $|\psi^{{\rm eff}}_s(T)\ra \approx 
|\psi_s(T)\ra$. The clock would have picked out the system state `at time T', as 
desired. However, recall that throughout this Section we have assumed that the system is 
bounded, in the sense that $E = 0$ was among the discrete spectrum of the 
operator $\hat{H}$. In this case even `good clocks' are not possible; 
$|f_c(a)|$ will generally return close to one for an infinite total range  
of $a$. In Section 5 we will discuss the `unbound' case, where $E = 0$ is among 
the continuous spectrum of $\hat{H}$. 
In that context non-repeating clock states will be plentiful, and such  
ambiguities are easily avoided.

Further understanding of $|\psi^{{\rm eff}}_s(T)\ra$ can be obtained by examining 
the spectra of $\hat{H}_s$ and $\hat{H}_c$. Suppose that equation (\ref{3.7}) is an exact 
equality, and that there are $N$ eigenvalues $E^s_{i}$ in the spectrum of  
$\hat{H}_s$ for which $E^c_i = - E^s_i$ is in the spectrum of $\hat{H}_c$. ($\{ E_i^s \}$ 
will be discrete, and will be finite whenever $\hat{H}_c$ and $\hat{H}_s$ are 
bounded below). Then $\hat{P}^{{\rm ph}}$ can be written in terms of the spectral 
projectors in $\clh_s$ and $\clh_c$ as
\be \hat{P}^{{\rm ph}} = \sum_{i=1}^{N} \hat{P}_{E^c_i} \hat{P}_{E^s_i} \label{34}\ee
Using this, $|\psi^{{\rm eff}}_s(T)\ra$ is given by:

\begin{eqnarray} |\psi^{{\rm eff}}_s(T)\ra & = \sum_{i=1}^{N}  
A_i \hat{P}_{E^s_i} |\psi_s\ra e^{- i E^s_i T}  \label{3.152}\\
\mbox{ where } \, A_i & = \la \psi_c | \hat{P}_{E^c_i}|\psi_c \ra \end{eqnarray}

	First consider the case when the spectrum of $\hat{H}_s$ and $-\hat{H}_c$ have only 
one eigenvalue in common. In this case $\hat{P}^{\rm ph} = \hat{P}_{E^c} \hat{P}_{E^s}$ does not entangle states in 
$\clh_s$ with states in $\clh_c$, so $|\psi^{{\rm eff}}_s(T)\ra$ is stationary, and 
$P(A \mbox{ when } T;\hat{\rho})$ is independent of $T$ for all $\hat{\rho}$ and all 
clocks $\hat{P}_T$. The CPI would not be viable for this 
choice of $\hat{H}_c$ and $\hat{H}_s$ (although this conclusion needn't apply if 
equation (\ref{3.7}) is not exact). If the eigenspaces for $E^s_i$ and $E^c_i$ are both 
degenerate, we could seek a better choice of $\hat{H}_c$ and $\hat{H}_s$ in 
equation (\ref{3.7}) which lifts this degeneracy, and provides the necessary 
correlations between system and clock. If, however, the Hamiltonian constraint has 
a unique solution (a possibility first suggested for quantum gravity by DeWitt \cite{DeWitt}), then this is not possible. For such a universe no Conditional 
Probability Interpretation would be possible in which equation (\ref{3.7}) was exact.

	Consider, on the other hand, that $N$ is large. If the clock state $|\psi_c \ra$ is chosen 
such that all $A_i$ are equal, then 
\be |\psi^{{\rm eff}}_s(T)\ra \propto |\psi^{{\rm relevant}}_s(T)\ra = \sum_{i=1}^{N} 
\hat{P}_{E^s_i} |\psi_s(T)\ra \ee
which is as close to $|\psi_s(T)\ra$ as is possible for this choice of $\hat{H}_s, \hat{H}_c$. For other choices of clock state, the effective system state $|\psi^{{\rm eff}}_s(T)\ra$ could bare little resemblance to 
$\psi_s(T)\ra$. Even then however, no violation of the Schr\"{o}dinger equation would be observed 
in $\clh_s$, since we still have $|\psi^{{\rm eff}}_s(T)\ra = e^{- i \hat{H}_s T} 
|\psi^{{\rm eff}}_s\ra$ (equation (\ref{3.10})). The only limitation is that the 
effective initial state $|\psi^{{\rm eff}}_s\ra$ is not the `chosen system state' 
$|\psi_s\ra$. An imperfect clock limits our ability to prepare a perfect `initial state', 
but does not affect the apparent evolution of this state.

\vs{.1}

	We now turn to the main objection to the Conditional Probability 
Interpretation, made by Kuchar in 1992 \cite{Kuchar} and recorded also in 
\cite{Page1}. To quote Kuchar \cite{Page1} ``You always apply the conditional 
probability formula to calculate the conditional probability of a projector at a 
single instant of an internal clock time. You never apply it to answering the 
fundamental DYNAMICAL question of the internal Schr\"{o}dinger interpretation, 
namely ``If one finds the particle at $Q'$ at the time $T'$, what is the probability 
of finding it at $Q''$ at the time $T''$?''. By your formula, that conditional 
probability differs from zero only if $T' = T''$ and $Q' = Q''$. In short, your 
interpretation prohibits the time to flow and the system to move!'' 

	This is an important objection, which Kuchar refers to as a ``reduction 
ad absurdum of the condition probability proposal''. Pages response (also recorded 
in \cite{Page1}) was ``In my viewpoint, only quantities at a single instant of 
time are directly accessible, and so one cannot directly test the two-time 
probability you discuss.'' We do not intend to defend this response here. It 
is the opinion of the author that, if Kuchar's claim were correct, it would 
indeed amount to a reduction ad absurdum of the CPI. However, lets investigate 
how this conclusion was reached. The formula used by Kuchar was (equation (13.19) 
of \cite{Kuchar}, which we have adjusted to our notation, and to states 
that are not necessarily pure)

\be P_{\rm Kuchar}(Q'' \mbox{ when } T''| Q' \mbox{ when } T' ; \hat{\rho}) = \frac{\Tr(\hat{P}_{Q' T'} \hat{P}_{Q'' T''} \hat{P}_{Q' T'} \hat{\rho}^{{\rm ph}})}{\Tr(\hat{P}_{Q' T'} \hat{\rho}^{{\rm ph}})} \label{kuchar} \ee

This is in disagreement with equations (\ref{twotime1}) - (\ref{twotime2}), and was 
never advocated by any proponent of the CPI. The first time equation (\ref{kuchar}) appeared 
was in Kuchar's criticism of it \cite{Kuchar}. It cannot be written in terms of physical 
measurement operators $M_{Q'' T''}$ and $M_{Q' T'}$, so it cannot be written as 
$P(Q'' T'' ; \hat{\rho}_{Q' T'})$ (in the sense of equation (\ref{3.1})) for any 
state $\hat{\rho}_{Q' T'}$. In classical terms, the numerator represents the 
proportion of the physical path on which $Q'', T'', Q', T'$ 
all occur simultaneously. It is not surprising therefore, that it is zero unless 
$Q'=Q''$ or that, for perfect clocks, (which Kuchar was referring to in the earlier quote) 
it is zero whenever $T' \neq T''$! 

	Lets investigate then, the predictions of equations 
(\ref{twotime1}) - (\ref{twotime2}). We consider the more general probability 
$P(\psi^{\rm out}_s \mbox{ when } T_2| \psi^{\rm in}_s \mbox{ when } T_1;\hat{\rho})$, 
where the projection $\hat{P}_{\psi^{\rm in}} = |\psi_s^{\rm in}\ra \la \psi_s^{\rm in}|$ 
on $\clh_s$ specifies the `initial system state', and $\hat{P}_T = |\psi_c(T)\ra \la 
\psi_c(T)|$ on $\clh_c$, just as in equation (\ref{new30}). The projection operator 
$P_{\psi^{\rm in}} P_{T_1}$ specifies a unique state 
$|\psi_s^{\rm in}\ra|\psi_c(T_1)\ra \equiv |\psi^{\rm in}, T_1 \ra$ in $\clh_{\rm aux}$ 
so we have:

\be \hat{P}_{\psi^{\rm in},T_1}^{{\rm ph}} \hat{\rho}^{{\rm ph}} \hat{P}_{\psi^{\rm in},T_1}^{{\rm ph}} \propto \hat{P}_{\psi^{\rm in},T_1}^{{\rm ph}} \ee
regardless of $\hat{\rho}^{\rm ph}$. Equation (\ref{twotime2}) gives:

\be P(\psi^{\rm out}_s \mbox{ when } T_2| \psi^{\rm in}_s \mbox{ when } T_1;\hat{\rho}) = \frac{|\la \psi^{\rm out}, T_2  | \hat{P}^{{\rm ph}} |\psi^{\rm in}, T_1 \ra|^2}{\sum_i |\la \psi^i, T_2  | \hat{P}^{{\rm ph}} |\psi^{\rm in}, T_1 \ra|^2} \label{41} \ee
where $i$ runs over a basis of $\clh_s$. Equation (\ref{13}) allows us to write:

\begin{eqnarray}
\hs{-2} \la \psi^{\rm out}, T_2  | \hat{P}^{{\rm ph}} |\psi^{\rm in}, T_1 \ra & = 
\lim_{\tau \rightarrow \infty} \frac{1}{2 \tau} \int_{-\tau}^{\tau} \, d {\rm a} \la \psi_c | e^{- i \hat{H}_c (a + T_1 - T_2)}| \psi_c\ra \la \psi^{\rm out}_s | e^{-i \hat{H}_s a} | \psi^{\rm in}_s \ra \label{3.19} \\
 & = \lim_{\tau \rightarrow \infty} \frac{1}{2 \tau} \int_{-\tau}^{\tau} \, d {\rm a} f_c(a) \la \psi^{\rm out}_s | e^{-i \hat{H}_s (a + T_2 - T_1)} | \psi^{\rm in}_s \ra \label{3.20} \\
 & = \la \psi^{\rm out}_s | e^{-i \hat{H}_s (T_2 - T_1)} | \psi^{\rm eff,in}_s \ra
\label{3.21} \end{eqnarray}

where the LHS is evaluated in $\clh_{\rm aux}$ and the RHS in $\clh_s$. The conditional probability is then given by:
\be P(\psi^{\rm out}_s \mbox{ when } T_2| \psi^{\rm in}_s \mbox{ when } T_1;\hat{\rho}) = \frac{|\la \psi^{\rm out}_s | e^{-i \hat{H}_s (T_2 - T_1)} | \psi^{\rm eff,in}_s \ra|^2}{\la \psi^{\rm eff,in}_s | \psi^{\rm eff,in}_s\ra} \label{45}\ee

It is an exact transition probability in $\clh_s$ between the out-state 
$|\psi^{\rm out}_s\ra$ and the effective in-state $| \psi^{\rm eff,in}_s\ra$. Again, 
the choice of clock state $|\psi_c\ra$ determines the choice of effective in-state 
$| \psi^{\rm eff,in}_s\ra$, which undergoes an `evolution' with respect to clock 
time that is indistinguishable from Schr\"{o}dinger evolution in $\clh_s$. It is 
the physical projection operator $\hat{P}^{\rm ph}$ in equation (\ref{3.19}) (absent 
in Kuchar's proposal (\ref{kuchar})) which imposes the correlations between clock 
and state that dictate this effective Schr\"{o}dinger evolution in $\clh_s$. 

	This shows that the CPI, at least in its refined form, is perfectly applicable 
to comparing more than one clock time, and that it makes appropriate predictions in 
such cases. We now turn our attention to systems which are `unbounded' in the sense 
that $E=0$ is among the continuous spectrum of the Hamiltonian.

\section{The Refined CPI for `Unbounded' systems.}

	When $E=0$ is among the continuous spectrum of $\hat{H}$ the `eigenstates' 
$|E,k\ra$ are generalized states. They are only $\delta$-function normaliseable:

\be \la E,k | E',k'\ra =  2 \pi \delta(E - E') \delta(k - k')  \label{4.1} \ee
where the index $k$ (possibly dependent on $E$, and possibly discrete) resolves 
the degeneracy of the relevant eigenspace (the factor $2 \pi$ is for later 
convenience). Hence, none of the physical states 
$|0,k\ra$ are actually in $\clh_{\rm aux}$ - they all have infinite norm! Accordingly, 
 equation (\ref{13}) gives zero in the limit for any $|\psi\ra 
\in \clh_{\rm aux}$. We could choose to continue regardless (as was advocated in 
the original treatments of the CPI - see pg 151 of \cite{Page4} for a discussion of 
this point), accepting that 
equations such as (\ref{essence1}) and (\ref{41}) will always be of the form $\frac{0}{0}$. 
Physically, this is not entirely unreasonable - it can be traced back to the fact that, 
if the physical path is infinite in length, 
then the proportion associated with any finite portion (such as our lifetime) will 
be zero, even though questions about that portion may be physically 
consistent. That said, 
the absence of a finite inner product on physical states, and the ill-defined nature of 
equations such as (\ref{13}), is at best disconcerting, and a more rigorous treatment is 
clearly desirable. This task has been considered by various authors \cite{Rieffel, Landsman,Higuchi,Mar94.1,Mar94.2,Mar95,Mar99,MarGi,Mar00,Ashtekar} under various names - the 
`induced' or `Rieffel induced' inner product, `Refined Algebraic Quantization', `Group 
Averaging' or the `Spectral Analysis' inner product. While research is 
ongoing on these topics (see \cite{Mar00} for a progress report) 
a substantial level of rigor has already been achieved, and the equivalence 
and uniqueness of these procedures has been established 
for typical cases \cite{Mar99,MarGi}. We will present only a brief outline of the induced 
Hilbert space here - the reader is referred elsewhere \cite{Mar99,MarGi,Mar00,Ashtekar} for 
more detail and rigor. 

	Loosely, the induced Hilbert space $\clh_{\rm ph}$ is constructed 
from the generalized states $|0,k\ra$ by replacing the inner product (\ref{4.1}) with 
the definition:

\be \la 0,k|0,k'\ra_{\rm ph} \equiv \delta(k - k') \ee
	which is equation (\ref{4.1}) `divided by $2 \pi \delta(0)$'. More generally, 
write two generalized eigenstates $|\phi\ra_{\rm ph}$, $|\psi\ra_{\rm ph}$ in terms 
of normaliseable states $|\phi_0\ra$, $|\psi_0\ra$ in $\clh_{\rm aux}$ through the action of 
the `operator' $\hat{P}^{\rm ph}$ (no longer strictly a projection on $\clh_{\rm aux}$):

\be |\phi\ra_{\rm ph} = \hat{P}^{\rm ph} |\phi_0\ra = \int {\rm d} E \, {\rm d} k |E,k\ra \delta(E)  \la E,k| \phi_0\ra \ee
Whereas $\la \hat{P}^{\rm ph} \psi_0|\hat{P}^{\rm ph} \phi_0\ra$ contains the infinite factor
 $2 \pi \delta(0)$, we can write the finite `induced' inner product on $\clh_{\rm ph}$ as:
\be \la \psi |\phi\ra_{\rm ph} \equiv \la \psi |\hat{P}^{\rm ph} |\phi\ra_{\rm aux} 
= \int {\rm d} E \delta(E) \int d k \la \psi_0 |E,k\ra \la E,k| \phi_0\ra \label{inprod1new}\ee
which is independent of the particular states $|\psi_0\ra$, $|\phi_0\ra$ chosen to 
represent $|\psi\ra_{\rm ph}$ and $|\phi\ra_{\rm ph}$. In many situations (see 
\cite{Ashtekar,Mar00,MarGi} for instance) $\hat{P}^{\rm ph}$ can also be written 
in terms of the `group averaging' procedure, as:

\be \hat{P}^{\rm ph} |\phi\ra = \int {\rm d} a \, e^{-i \hat{H} a} |\phi\ra \label{4.00}\ee
and equation (\ref{inprod1new}) becomes\footnote{When necessary, equation (\ref{inprod3}) 
is understood as $\lim_{\epsilon \rightarrow 0^+} \int {\rm d} a g_{\epsilon}(a) \la \phi 
| e^{- i \hat{H} a}  | \psi \ra_{\rm aux}$, where $g_{\epsilon}(a)$ is a positive integrable function for which $\lim_{\epsilon \rightarrow 0^+} g_{\epsilon}(a) = 1$ for all $a$.}:
\be \la \phi | \psi \ra_{\rm ph}  = \int {\rm d} a \la \phi | e^{- i \hat{H} a}  | \psi \ra_{\rm aux} \label{inprod3} \ee

Equation (\ref{4.00}) can be compared with equation (\ref{13}). It projects out the 
zero-frequency component of the Schr\"{o}dinger-evolved state. To gain more familiarity 
with equation (\ref{4.00}) consider $\clh_{\rm aux} = \clh_c \otimes \clh_s$ and 
consider system and clock states contained entirely within the continuous spectrum of 
$\clh_s$, $\clh_c$ respectively. Then we can decompose the Schr\"{o}dinger states 
$|\psi_s(a)\ra = e^{- i \hat{H}_s a} |\psi_s\ra$ and $|\psi_c(a)\ra = e^{- i \hat{H}_c a} 
|\psi_c\ra$ in terms of their frequencies as:

\be |\psi_s(a)\ra = \int \frac{{\rm d} E}{2 \pi} \, \tilde{\psi}_s(E) e^{- i E a} \hs{1} |\psi_c(a)\ra = \int \frac{{\rm d} E}{2 \pi} \, \tilde{\psi}_c(E) e^{- i E a} \ee
In terms of these we find:

\begin{eqnarray} 
\hat{P}^{\rm ph} |\psi_c\ra |\psi_s\ra & = \int {\rm d} a \frac{{\rm d} E}{2 \pi} \, \frac{{\rm d} E'}{2 \pi} \tilde{\psi}_c(E) \tilde{\psi}_s(E') e^{-i (E + E') a} \\
& = \int \frac{{\rm d} E}{2 \pi} \tilde{\psi}_c(E) \tilde{\psi}_s(-E) \end{eqnarray}
which is the continuum equivalent of equation (\ref{34}).

Given a suitable projection operator $\hat{P}_A$ on $\clh_{\rm aux}$, we 
can define $\hat{P}_A^{\rm ph}$ on $\clh_{\rm ph}$ by:
\begin{eqnarray} 
\hs{-1.5} \la \psi |\hat{P}_A^{\rm ph} | \phi\ra_{\rm ph} & \equiv \la \psi_0 
| \hat{P}^{\rm ph} \hat{P}_A \hat{P}^{\rm ph} | \phi_0 \ra_{\rm aux} \label{op1} \\
 & = \int {\rm d} E {\rm d} E' \delta(E) \delta(E') \int {\rm d} k {\rm d} k' 
\, \la \psi_0| E k \ra \la E k| \hat{P}_A | E' k' \ra_{\rm aux} \la E' k'| 
\phi_0 \ra \label{op2} \\
 & = \int {\rm d} a {\rm d} a' \la \psi_0 |e^{-i \hat{H} a} 
\hat{P}_A e^{-i \hat{H} a'} |\phi_0\ra_{\rm aux} \label{op3}\end{eqnarray}
	where $\hat{P}_A$ is `suitable' if equations (\ref{op2}) and (\ref{op3}) 
converge for all $|\phi_0\ra$, $|\psi_0\ra$ in $\clh_{\rm aux}$ (or in some 
suitably large subspace $\Phi$ of $\clh_{\rm aux}$ \cite{Mar99}). This 
requirement on $\hat{P}_A$ excludes the possibility that $\hat{P}_A$ already 
commutes with the constraint. (It corresponds physically to the requirement that 
$\hat{P}_T$ project onto a `finite portion of the physical path'.) If $[\hat{P}_A,\hat{H}] 
= 0$, then $\la E,k|\hat{P}_A|E',k'\ra_{\rm aux}$ contains a factor of $\delta(E-E')$ and 
(\ref{op2}) becomes ill-defined. An alternative way to induce operators on $\clh_{\rm ph}$ 
is to consider {\it only} operators which commute with $\hat{H}$ and to define 
$\hat{P}_A^{\rm ph}$ by omitting one of the factors of $\hat{P}^{\rm ph}$ in (\ref{op1}). 
That is effectively the strategy in \cite{Mar99} for instance. That strategy 
is consistent with the strategy above, in the sense that, given an operator 
$\hat{P}_A$ satisfying the requirement above, we can define an 
operator $\hat{P}_A^{\rm new}$ on $\clh_{\rm aux}$ by
\be \hs{-1.5} \la \psi_0 | \hat{P}^{\rm new}_A | \phi_0 \ra_{\rm aux} = \int {\rm d} 
E {\rm d} E' \delta(E - E') {\rm d} k {\rm d} k' \, \la \psi_0| E k \ra \la E k| 
\hat{P}_A | E' k' \ra_{\rm aux} \la E' k'| \phi_0 \ra \label{59} \ee
This satisfies $[\hat{H},\hat{P}_A^{\rm new}] = 0$, making it suitable for the procedure 
described in \cite{Mar99}. That procedure applied to $\hat{P}_A^{\rm new}$ then leads 
to the operator $\hat{P}_A^{\rm ph}$ on $\clh_{\rm ph}$ as defined above. The definition 
in equations (\ref{op2}), (\ref{op3}) above 
is more suitable for our purposes, since we wish to construct physical 
operators $\hat{P}_T^{\rm ph}$ from clock projection operators $\hat{P}_T$ 
satisfying $[\hat{P}_T,\hat{H}] \neq 0$.

	Operators such as $\hat{P}^{\rm ph}_{B C}$ are obtained by replacing 
$\hat{P}_A$ in equations (\ref{op1}) - (\ref{op3}) with $\hat{P}_B \hat{P}_C$ while 
products $\hat{P}_B^{\rm ph} \hat{P}_C^{\rm ph}$ of physical projection operators 
can be obtained by replacing $\hat{P}_A$ in equations (\ref{op1}) - (\ref{op3}) with 
$\hat{P}_B \hat{P}^{\rm ph} \hat{P}_C$, to obtain:

\begin{eqnarray} 
\la \psi |\hat{P}_B^{\rm ph} \hat{P}_C^{\rm ph} | \phi\ra_{\rm ph} & \equiv 
\la \psi_0 | \hat{P}^{\rm ph} \hat{P}_B \hat{P}^{\rm ph} \hat{P}_C \hat{P}^{\rm ph} 
| \phi_0 \ra_{\rm aux} \label{op4} \\
 & = \int {\rm d} a {\rm d} a' {\rm d} a'' \la \psi_0 
|e^{-i \hat{H} a} \hat{P}_B e^{-i \hat{H} a'} \hat{P}_C e^{-i \hat{H} a''} 
|\phi_0\ra_{\rm aux} \label{op5}\end{eqnarray}

	These equations extend to density operators in the obvious way:


\begin{eqnarray}
\Tr_{\rm ph}(\hat{\rho}) & = \Tr_{\rm aux}(\hat{P}^{\rm ph} \hat{\rho_0}) = \int {\rm d} a \Tr_{\rm aux}(e^{-i \hat{H} a} \hat{\rho}) \\
\Tr_{\rm ph}(\hat{P}_A^{\rm ph} \hat{\rho}) & = \Tr_{\rm aux}(\hat{P}^{\rm ph} \hat{P}_A \hat{P}^{\rm ph} \hat{\rho}_0) = \int {\rm d} a {\rm d} a' \Tr_{\rm aux}(e^{-i \hat{H} a} \hat{P}_A e^{-i \hat{H} a'} \hat{\rho}) \end{eqnarray}

	The formulae of Section 3 (with $\hat{P}^{\rm ph} \hat{P}^{\rm ph}$ identified 
with $\hat{P}^{\rm ph}$ throughout) now carry over almost unchanged to the unbounded 
case. For instance:
\be P(A \mbox{ when } B;\hat{\rho}) = \frac{\Tr_{{\rm aux}}(\hat{P}^{\rm ph} \hat{P}_{A} \hat{P}_{B} \hat{P}^{\rm ph} \hat{\rho}_0)}{\Tr_{{\rm aux}}(\hat{P}^{\rm ph} \hat{P}_{B} \hat{P}^{\rm ph} \hat{\rho}_0)} \label{essence1new} \ee

The derivation of equation(\ref{3.11}) remains unchanged. If we assume equations 
(\ref{new30}) and (\ref{new31}), then we are again lead to equation (\ref{3.15}), with 
the effective system state $|\psi_s^{\rm eff}(T)\ra$ in $\clh_s$ given by:

\be |\psi^{{\rm eff}}_s(T) \ra = \int {\rm d} a \, f_c(a) |\psi_s(T + a)\ra = 
e^{-i \hat{H} T}|\psi^{{\rm eff}}_s \ra  \label{4.10} \ee
which is just equation (\ref{3.10}) without the factor $\frac{1}{2 \tau}$. It is 
generally straightforward to find a clock state $|\psi_c\ra$ such that 
$|f_c(a)| = |\la \psi_c | e^{-i \hat{H} a}|\psi_c \ra|$ is integrable. Such a clock can be said to `read zero for a finite Schr\"{o}dinger time'. This is consistent 
with the `good clock requirement', which requires also that $|f_c(a)|$ be sharply peaked 
about $a=0$. It ensures that $\hat{P}_T$ is `suitable' in the sense of equations (\ref{op2}) and (\ref{op3}), and hence ensures that 
$|\psi^{{\rm eff}}_s \ra$ has finite norm in $\clh_s$. 

	Equation (\ref{41}) is also unchanged, and leads to equation (\ref{45}) 
just as in the `bounded' case. The LHS of equation (\ref{3.19}), which is evaluated 
in $\clh_{\rm aux}$, is by definition the induced inner product $\la \psi^{\rm out}, 
T_2 |\psi^{\rm in}, T_1 \ra_{\rm ph}$ between $|\psi^{\rm in}, T_1 \ra_{\rm ph}$ and 
$|\psi^{\rm in}, T_1 \ra_{\rm ph}$. Equation (\ref{3.21}) shows that this equates to 
the S-Matrix element $\la \psi^{\rm out}_s | e^{-i \hat{H}_s (T_2 - T_1)} | 
\psi^{\rm eff,in}_s \ra$ in $\clh_s$ between the system states 
$ | \psi^{\rm eff,in}_s \ra$ and $| \psi^{\rm out}_s \ra$.

	The simplest application of this construction is to Parametrized 
Particle Dynamics, with $\hat{H}_s = \tilde{H}(\hat{p}_i,\hat{x}^i)$, and 
$\hat{H}_c = \hat{p}_t = -i \frac{\partial}{\partial t}$. The system Hilbert 
space $\clh_s$ is then the standard Hilbert space of non-relativistic quantum 
mechanics, while the clock Hilbert space $\clh_c$ contains square integrable 
functions of $t$. The clock function $f_c(a)$ is given by:

\be f_c(a) = \int {\rm d} t' \psi_c^*(t') \psi_c(t'-a) \ee
where $\psi_c(t) = \la t | \psi_c \ra$ is the initial clock state in `position 
representation' (in $\clh_c$). In the ideal clock limit $\psi_c(t) \rightarrow \delta(t)$, 
so that $f_c(a) \rightarrow \delta(a)$ and equation (\ref{4.10}) gives 
$|\psi^{{\rm eff}}_s \ra = |\psi_s \ra$. Conventional quantum mechanics is exactly 
retrieved in this limit.

	We have now shown how the CPI can be applied successfully whenever $E=0$ is 
among the continuous or the discrete spectrum of the Hamiltonian $\hat{H}$. These 
cases are of course not exhaustive - The $E=0$ eigenstates of the Hydrogen atom being 
an example that is neither of these cases. Constructing an `induced inner product' for  
these more difficult cases is considered in \cite{Mar99}. We now consider 
briefly how the CPI connects with other `timeless interpretations' of the Hamiltonian 
constraint. At the time of the original debate about the CPI \cite{Isham,Kuchar} there 
were two other prominent attempts at understanding the Hamiltonian constraint - Hartle's 
`consistent histories' approach \cite{Hartle1,Hartle2}, and Rovelli's `Evolving Constants 
of Motion' \cite{Rov1,Rov2,Rov3} (see also \cite{Hajicek,Hart2} for early criticisms). 
The consistent histories approach has been extensively developed since then 
\cite{Hart,Halliwell,HaTh,HaMa,Whelan}, while the work of Marolf 
\cite{Mar94.1,Mar94.2,Mar95} refined and developed the `evolving 
constants' approach. Marolf's construction is in many ways similar to the refined CPI 
presented in this Section. The induced inner product is used to construct operators 
on the induced Hilbert space; the steps from $\omega \rightarrow \Omega \rightarrow 
\Omega_{phys}$ in \cite{Mar95} for instance directly parallel the construction 
$\hat{P}_A \rightarrow \hat{P}_A^{\rm new} \rightarrow \hat{P}_A^{\rm phys}$ described 
after equation (\ref{59}). However, the operators used to `keep time' in \cite{Mar94.1,Mar95} 
are different to those used here. They choose a configuration space variable $Q$ and seek 
answers to the question ``what is the value of $A$ when $Q = \tau$?''. This involves 
constructing a time projection operator from the spectral projections of the 
Hamiltonian-evolved operator $\hat{Q}(t)$, while the integral over $Q$ 
plays a role analogous to that of $\clh_c$. This agrees with our construction for the case of 
Parametrized Particle Dynamics, but is more restrictive in general.

	We have considered the role of clock projection operators $\hat{P}_T$ 
throughout this paper, in order to describe how equation (\ref{3.3}) can be 
used to answer questions of the form ``What is the probability of $A$ {\it when } 
the clock reads $T$'' and to describe how, in conjunction with standard logical 
operations such as {\it and} and {\it given} it can be used to tackle more general 
questions, such as the two-time probability described in equation (\ref{twotime1}). However,
 the formalism is of course more general than this, and needn't be restricted to 
questions involving specified `clock times'. A projection $\hat{P}_V$ onto a region 
$V$ of configuration space could be used in equation (\ref{3.1}) for instance, to 
associate a probability with a given region of phase space, answering the quantum 
equivalent of the question ``what proportion of the observers worldline is in $V$''. The 
most general probability that can be constructed from the rules described in Section 3 
is of the form:

\begin{eqnarray} P_{\balpha_0,A} & = \frac{\Tr(\hat{C}_{\balpha_0}^{\dagger} \hat{C}_{\balpha_0} \hat{\rho})}{\sum_{\balpha \in A} \Tr(\hat{C}_{\balpha_0}^{\dagger} \hat{C}_{\balpha_0} \hat{\rho})} \\
\hs{-1.5} \mbox{ where } \hs{.5} \hat{C}_{\balpha} & = \hat{M}_{\alpha_1} \hat{M}_{\alpha_2} \dots \end{eqnarray}
	and the operators $\hat{M}_{\alpha_i}$ are physical measurement 
operators. Although the individual operators $\hat{C}_{\balpha}$ do not commute 
with the constraint (and are not hermitian), $\hat{C}_{\balpha}^{\dagger} \hat{C}_{\bbeta}$ 
commutes with the constraint for any $\balpha,\bbeta$. The operators $\hat{C}_{\balpha}$ 
can be identified with the class operators of the consistent histories approach 
\cite{Hart,Hartle1,Hartle2}, allowing connection to be made between the CPI and the 
consistent histories approach. (We haven't considered the decoherence functional 
$\Tr(\hat{C}_{\balpha_0}^{\dagger} \hat{C}_{\balpha_1} \hat{\rho})$ here, although 
we acknowledge that in specific cases this should be considered before treating the 
`histories' $\balpha$ as a family of decoherent alternatives.) Decoherent histories 
approaches have been developed based on the induced 
inner product \cite{HaTh,Halliwell} and on the Klein Gordon inner product 
\cite{Hart,Whelan}. The detailed connection between the CPI and these specific 
approaches is yet to be investigated. 

	Another point worth addressing is the fact that in practice the `system' 
$\clh_s$ and the `clock' $\clh_c$ do not together constitute the whole universe. Consider 
for instance, replacing equation (\ref{3.7}) with:

\be \hat{H}_{tot} \approx \hat{H}_{s} + \hat{H}_{c} + \hat{H}_{\rm rest} \ee

where $[\hat{H}_{\rm rest},\hat{H}_s] = 0 = [\hat{H}_{\rm rest},\hat{H}_c]$. This 
results in the minor change:
\be P(A \mbox{ when } T) = \Tr_{{\rm s}}(\hat{P}_{A} e^{-i \hat{H}_s T} \hat{\rho}^{{\rm s}}e^{i \hat{H}_s T}) \hs{.3} \mbox{ where } \hs{.5} \hat{\rho}^{{\rm s}} \equiv \frac{\Tr_{{\rm c,rest}}( \hat{P}_{0} \hat{\rho}^{{\rm ph}})}{\Tr_{{\rm aux}}(\hat{P}_{0} \hat{\rho}^{{\rm ph}})} \label{3.16} \ee

The system still behaves {\it as if} $\hat{\rho}_s$ is Schr\"{o}dinger evolved 
in $\clh_s$ with the only difference being the trace over $\clh_{\rm rest}$ in 
(\ref{3.16}). We might ask whether this trace can destroy the correlations between 
clock and system that are necessary for the clock to `keep time'. Fortunately this 
is not so. Consider for instance the simple case described in equation (\ref{new31}), 
with $\hat{\rho}^{{\rm ph}} = \hat{P}^{{\rm ph}} |\psi_s\ra |\psi_c\ra \la 
\psi_c|\la \psi_s| \hat{P}^{{\rm ph}}$. The `effective initial density operator' 
$\hat{\rho}^s$ in equation (\ref{3.16}) is then given by:
\be \hat{\rho}_s \propto \int {\rm d} a {\rm d} a' f(a) f^*(a') \Tr_{\rm rest}(e^{-i \hat{H}_r (a - a')}) |\psi_s(a) \ra \la \psi_s(a')| \label{3.17} \ee

If there is no trace over $\clh_{\rm rest}$ then equation (\ref{3.17}) factorizes, 
to give $\hat{\rho}_s \propto |\psi_s^{\rm eff} \ra \la \psi_s^{\rm eff} |$ as in 
equation (\ref{3.15}). If on the other hand $\clh_{\rm rest}$ is very large as is 
typical in everyday systems, then we would expect in general that 
$\Tr_{\rm rest}(e^{-i \hat{H}_r (a - a')})$ would be sharply peaked about $a=a'$. Then  
$\hat{\rho}_s$ is given, up to an overall factor, by:

\be \hat{\rho}_s \approx \int {\rm d} a |f(a)|^2 |\psi_s(a) \ra \la \psi_s(a)| \label{3.18} \ee

This state is no longer pure, but is still a good approximation of the 
initial system state whenever $f(a)$ is sharply peaked around $a=0$. 
The trace over $\clh_{\rm rest}$ has not destroyed the correlations necessary 
for the clock to `keep time', because these correlations are established in 
$\hat{P}_{0} \hat{\rho}^{{\rm ph}}$. Since the trace is taken over 
$\hat{P}_{0} \hat{\rho}^{{\rm ph}}$ and not over $\hat{\rho}^{{\rm ph}}$ itself, 
then it does not affect time-keeping. For `repeating' clocks, equation (\ref{3.18}) 
is in some ways more natural than equation (\ref{3.15}). Suppose for instance that 
$|f(a)|$ was sharply peaked about two different values of $a$; such a clock would `read 
zero on two occasions'. Then equation (\ref{3.18}) registers these occasions as 
classical alternatives, whereas equation (\ref{3.15}) would include interference 
terms between the two occasions.

\section{Conclusion}

	The Conditional Probability Interpretation (CPI) has been reviewed, and 
minor refinements proposed. We have explained how, at least in it's refined form, 
the CPI is capable of answering various past criticisms \cite{Kuchar,UW,Isham}. In 
particular, questions involving more than one clock time were described in detail, 
resolving the problems raised in \Kuchar's ``reduction ad absurdum'' 
\cite{Kuchar,Page1}. In the case of Parametrized Particle Dynamics, conventional 
quantum mechanics was exactly recovered in the ideal clock limit. Situations 
were addressed where $E=0$ was among the continuous spectrum of the Hamiltonian. The 
induced inner product \cite{Rieffel,Landsman,Higuchi,Mar94.1,Mar94.2,Mar99} was used 
to construct the physical Hilbert space $\clh_{\rm ph}$ from the generalized 
eigenvectors in (the topological dual of) $\clh_{\rm aux}$. This allowed the CPI 
to be applied to these `continuous-spectrum' cases in a more rigorous fashion than 
that described previously \cite{Page1,Page2,Page3,Page4}. This induced construction 
was described in outline only here - more rigor in that construction is possible, 
desirable, and current \cite{Mar99,MarGi,Mar00}. A useful feature of this induced 
construction, particularly in conjunction with group averaging techniques 
\cite{Ashtekar,MarGi,Mar00} is that, once the existence of $\clh_{\rm ph}$ has 
been established, we can proceed to work entirely in $\clh_{\rm aux}$. We needn't 
solve the eigenvalue equation directly, since this is achieved implicitly through 
the integral representation of $\hat{P}^{\rm ph}$. Nor must we find operators 
$\hat{P}_A$ which have clear physical meaning {\it and} which directly commute 
the Hamiltonian. The definition (\ref{inprod3}) allows us to define such physical 
operators implicitly, and helps to suggest their physical interpretation. The 
induced Hilbert space and group averaging techniques are also prominent in 
current approaches to quantum gravity \cite{Ashtekar,Ashtekar2,Rovelli,Rovelli2,Thiemann}, 
while the auxiliary Hilbert space plays a similar role to that played here, suggesting 
that the CPI could be ideally suited for interpreting the role of time in those theories.

	The Conditional Probability Interpretation does not require a `global time 
coordinate' on the configuration space, or a `time operator' $\hat{T}$ on the total 
Hilbert space. The choice of clock projection operators $\hat{P}_T$ can be tailored 
to the subsystem $\clh_s$ we are interested in, just as in ordinary physics, where 
different time-keeping devices (atomic clocks, wrist-watches, planetary motions, or 
the universes scale factor $a$) are appropriate for measuring change in different 
situations. Neither the physical Hilbert space $\clh_{\rm ph}$ nor the auxiliary 
Hilbert space $\clh_{\rm aux}$ depend in any way on our choice of time-keeping 
device. This flexibility in the choice of time-keeping device allows the CPI to be 
fully compatible with situations where no unique `time coordinate' exists, thus 
avoiding the ``multiple choice'' and ``global time'' problems described in 
\cite{Kuchar}. Indeed, equation (\ref{3.3}) allows us to give meaning to the 
statement `A {\it when } B' even when neither A nor B are `clocks' (in the 
sense of equation (\ref{3.8})). The apparent unitary evolution of $\hat{\rho}_s(T)$ 
in $\clh_s$ stems directly from equation (\ref{3.8}). The Hamiltonian constraint 
then forces the correlations between clock and system that ensure the effective 
Schr\"{o}dinger-evolution in $\clh_s$. This is why, as observers within a constrained 
system, the correlations between ourselves and the world around us allow us to 
observe changes in that world, and why isolated subsystems change with respect to `clock 
time' in a way that is compatible with the time-dependent Schr\"{o}dinger equation.

\end{document}